\documentstyle[12pt]{article}
\newcommand{\cl}{\centerline}
\begin{document}
\begin{titlepage}


\vfill
\begin{center}
{\large{{\bf Characterization of fractional-quantum-Hall-effect
quasiparticles}}\\}\par
\vskip 1.5cm
{Alfred S. Goldhaber$^{a,b,}$\footnote{goldhab@insti.physics.sunysb.edu}
and J.K. Jain $^{a,}$\footnote{jain@insti.physics.sunysb.edu}\\}
\vskip 1.0cm
{\it {$^{a}$Department of Physics and $^{b}$Institute for Theoretical Physics,
State University of New York,\\}
{Stony Brook, New York 11794-3800, USA.\\}}
\end{center}
\vskip 0.75cm
\cl{\today}
\noindent
\vskip 0.75 cm
\vskip 1.5cm
\cl{\bf Abstract}

Composite fermions in a partially filled quasi-Landau level
may be viewed as quasielectrons of the underlying  fractional quantum
Hall state, suggesting that a quasielectron is
simply a dressed electron, as often is true in other
interacting electron systems, and as a result
has the same intrinsic charge and
exchange statistics as an electron.  This paper
discusses how this result is reconciled with the earlier picture in
which quasiparticles are viewed as fractionally-charged
fractional-statistics ``solitons". While the two approaches provide
the same answers for the long-range interactions
between the quasiparticles, the dressed-electron description is more
conventional and unifies
the view of quasiparticle dynamics in and beyond the
fractional quantum Hall regime.
\vfill \end{titlepage}

\newpage

{\bf INTRODUCTION}

An intriguing feature of the fractional quantum Hall effect (FQHE)
\cite {Tsui} is the
existence of two theoretical perspectives, one formulated in terms of
fractionally-charged fractional-statistics quasiparticles \cite
{Laughlin83,Halperin84,Haldane83}, and the
other in terms of charge  $-e$ fermions, called composite
fermions \cite {Jain89}. Such a circumstance occurs often in
physics, and experience shows that we can gain insight by exploring
all consistent perspectives, since for particular purposes one or
another may be advantageous. The purpose of this note is to
analyze the different ways in which the two approaches mentioned above
obtain the same results for
a set of quantities characterizing the topological properties of a
given incompressible FQHE state. The considerations below are simple
and straightforward, but to the best of our knowledge  have not been
spelled out clearly in the literature.

In both descriptions, Laughlin's elegant and successful wave function
has played a crucial role:  At filling factor $\nu=1/(2m+1)$,
where $m$ is an integer, the Laughlin ground state \cite {Laughlin83} is
\begin{equation}
\chi_{_{L}}=\prod_{j<k}(z_{j}-z_{k})^{2m+1}\exp[-\frac{1}{4}\sum_{i}|z_{i}|^2],
\label{lwf}
\end{equation}
where $z_{j}=x_{j}-iy_{j}$ denotes the position of the $j$th electron.
The picture which arises from this wave function
is one in which the quasiparticles
(we use the word `quasiparticle' to denote either a `quasielectron' or
a `quasihole')
are considered as topological solitons, carrying fractional
electric charge $|e^*| = e/(2m+1)$  \cite {Laughlin83}.
Further, they are anyons, i.e., when one quasiparticle moves halfway around
another, this produces a fractional `statistics phase' $\pi\theta$,
where $\theta=1/(2m+1)$ \cite {Halperin84,ASW}.
To describe filling factors other than $\nu=1/(2m+1)$,
the quasiparticle-hierarchy  (QPH) approach starts with a `parent' state,
and builds Laughlin states out of its fractionally-charged
fractional-statistics quasiparticles to produce `daughter'
states \cite {Halperin84,Haldane83}. An iteration of this procedure
obtains the possibility of FQHE at all odd-denominator fractions
starting from the Laughlin states at $\nu=1/(2m+1)$. We consider here
only the principal fractions $n/(2mn+1)$. The charge of the
quasiparticle of this state is given by \cite
{Laughlin83,Halperin84,Haldane83}
\begin{equation}
|e^*|=\frac{e}{2mn+1}\;\;,
\end{equation}
and the ``anyon statistics" by  \cite {Halperin84}
\begin{equation}
\theta=\frac{2m(n-1)+1}{2mn+1}\;,
\end{equation}
defined so that a complete loop of one quasiparticle around another
produces a phase factor $e^{i2\pi \theta}$.
(This definition of statistics has to do with the
dynamical phase acquired by the wave function as quasiparticles
wind around each other. We call it ``anyon statistics" to distinguish
it from the usual kinematical exchange statistics, though it is a
common practice to refer to the net phase simply as `statistics'
without qualification.)

Composite fermions (CF) are electrons carrying an even number of
vortices of the many-particle wave function \cite {Jain89}. In the CF
theory, the strongly correlated liquid of interacting electrons at
$\nu$ is mapped on to a weakly interacting gas of composite
fermions at $\nu^*$.  The zero-order wave
function $\chi_{\nu}$ of interacting electrons at $\nu$ is given by
\begin{equation}
\chi_{\nu}=\prod_{j<k}(z_{j}-z_{k})^{2m}\Phi_{\nu^*}
\label{cfwf}
\end{equation}
where $\Phi_{\nu^*}$ is a Slater determinant for non-interacting
electrons at $\nu^*$. The Jastrow factor
$\prod_{j<k}(z_{j}-z_{k})^{2m}$ generates $2m$ vortices in $\chi_{\nu}$
in the relative coordinate of each
electron pair, converting the electrons into composite fermions.
The LL's of electrons in
$\Phi$ translate into the quasi-LL's of composite fermions (after
multiplication by the Jastrow factor). The  right hand side of
Eq.~(\ref{cfwf}) is interpreted as composite
fermions at effective filling factor $\nu^*$, which is related to the
electron filling factor $\nu$ by
\begin{equation}
\nu=\frac{\nu^*}{2m\nu^*+1}\;\;.
\label{ff}
\end{equation}
In particular, the state with $n$ filled quasi-LL's of composite
fermions corresponds to the incompressible $\nu=n/(2mn+1)$ FQHE state.
The lowest Landau level projections of these wave functions are known
to provide a very good description of the actual eigenstates
for systems with small numbers of electrons \cite
{Dev}.  For the special case $\nu^*=1$, this ground state wave function
becomes the original Laughlin wave function \cite {Laughlin83}.
For the purposes of this paper, it will be assumed that various
long-distance properties derived with the help of these wave
functions are exact.

A related theoretical approach to describing the FQHE states
is the introduction of Chern-Simons (C-S) interactions supplementing the
electromagnetic interactions.  These interactions may be chosen either
so that the quasiparticles are fermions \cite {Lopez}, in which case the
description is equivalent to the CF picture, or so that the quasiparticles
are bosons \cite {GM,ZHK}, in which case the Laughlin states are
described at the first step, and the other FQHE states are
obtained in analogy with the Haldane-Halperin hierarchy
scheme. Thus the comparisons made
here apply also to the C-S approaches even though we do not
discuss them explicitly.

{\bf QUASIPARTICLES AS DRESSED ELECTRONS}

In the CF theory, it is natural to identify the composite fermions
in the topmost partially filled quasi-LL as the objects analogous to
the quasielectrons of the QPH scheme. (Equivalently,
the holes in the topmost partially filled
LL of $\Phi_{\nu^*}$ map into the quasiholes of the FQHE state at
$\nu$. For simplicity, we restrict our attention to
quasielectrons.)  Since composite fermions are simply
dressed electrons, this identification has two immediate consequences:
(i) Quasielectrons have intrinsic charge $-e$. (ii) Their exchange produces
a sign $-1$, i.e., they are fermions. These quantities are independent
of the specific FQHE state in question.
This appears to be in fundamental conflict with the
QPH theory,  which ascribes a FQHE-state-specific intrinsic
charge $-e^*$ and fractional statistics $\theta$ to quasiparticles.
The objective of this paper is to show that despite the apparent
dissimilarity, a mean-field treatment of composite fermions
provides the same answers for various topological
phases as the QPH scheme.

{\bf ``LOCAL CHARGE" OF QUASIPARTICLES}

A crucial piece of information in the following is the concept of
``local charge". While the intrinsic charge of a CF-quasielectron is
$-e$, it is screened by the CF medium, which produces
a ``correlation hole" around each quasielectron, and the
sum of intrinsic charge of the quasielectron and the charge of the
correlation hole, {\em measured relative to the charge density
corresponding to the background FQHE state}, is only a
fraction of $-e$. We call this charge the ``local charge" of the
quasielectron, which can also be defined as the charge in a sufficiently
large area containing a quasielectron minus the charge
in the same area if it contained no quasielectron. It obviously
depends on the screening properties of the background CF state,
described by the microscopic wave functions of Eq.~(\ref{cfwf}).
There are several ways of obtaining the local charge \cite
{Laughlin83,Halperin84}. We briefly repeat here a derivation that uses a
counting argument \cite {Jaincomm}.
Start with the $n/(2mn+1)$ FQHE state confined to a disk
of a given radius, which fixes the largest allowed power of $z_{j}$.
Now add an {\em electron} to this system while insisting that
the size of the system not change. The product in the Jastrow factor
in Eq.~(\ref{cfwf}) now goes from 1 to
$N+1$, which increases the largest power of $z_{j}$ in the Jastrow
factor by $2m$. Therefore, in order to stay within the allotted area,
the largest power of $z_{j}$ in $\Phi$ must be
reduced by $2m$, which requires taking $2m$ electrons from the boundary of
each LL and putting them in the interior of the $(n+1)^{st}$ LL.
Including the new
electron, the $(n+1)^{st}$ LL of $\Phi$ now has $2mn+1$ electrons, i.e.,
the $(n+1)^{st}$ quasi-LL of $\chi$ now has $2mn+1$ composite fermions
(or quasielectrons). Since a charge $-e$ was added, each quasielectron
has a local charge of
\begin{equation}
-e^*=\frac{-e}{2mn+1}\;\;,
\end{equation}
which is the same as the intrinsic charge of a quasielectron in the
QPH description.  Thus, as the added electron
gets screened by the medium into a quasielectron of local
charge $-e^*$, $2mn$ additional composite fermions are excited out of
the vacuum.  The above argument remains valid even away from a FQHE
state, when there are several quasiparticles. The
local charge of the quasielectron
is still given by $-e^*=-e/(2mn+1)$, where $n$ is
the number of quasi-LL's whose edges are occupied.
It is worth emphasizing that it is a unique property of an
incompressible CF state
that the quasiparticle charge is only partially screened (usually,
it is either fully screened or not at all),
precisely so as to give a
simple fractional value for the local charge.

{\bf TOPOLOGICAL PHASES}

In the CF picture, the quasielectron has intrinsic charge $-e$,
local charge $-e^*$, and obeys fermion exchange statistics.
We will be concerned with the phase acquired when a composite fermion
(which may be a quasielectron) is taken (counterclockwise) around
a closed loop enclosing an area $A$. It has two contributions:
one is the usual Aharonov-Bohm (AB) phase, $2\pi BA/\phi_{0}$,
due to a charge $-e$  taken around the loop, where
$\phi_{0}=hc/e$ is the quantum of flux, and the other contribution
comes from the vortices on other composite fermions inside the
loop. The total phase is
\begin{equation}
2\pi \frac{BA}{\phi_{0}}-4m\pi K,
\label{phase}
\end{equation}
where $K$ is the number of composite fermions enclosed in this loop.
The second term is a direct consequence of the binding of vortices to
electrons, i.e., the formation of composite fermions. This expression
for the phase associated with a closed loop is independent of the
details of the FQHE state in question.

In a mean-field approximation, similar to the one proposed
by Laughlin \cite {Laughlin88} in his theory of anyons, $K$ is
replaced by the average number of composite fermions inside the loop.
For the special case of uniform particle density,
$<K>=A\rho$, where $\rho$ is the particle  density per unit area, and the
average phase associated with a closed loop is given by
\begin{equation}
\frac{2\pi A}{\phi_{0}}(B-2m\rho\phi_{0})\;\;.
\label{mfphase}
\end{equation}
Therefore, in a mean-field sense the composite fermions move
as if they were in an effective field
\begin{equation}
B^*=B-2m\phi_{0}\rho\;\;.
\end{equation}
This equation is identical to Eq.~(\ref{ff}) with
$\nu^*=\rho\phi_{0}/B^*$ and $\nu=\rho\phi_{0}/B$.
In particular, when the effective filling factor of composite fermions
is integer ($\nu^*=n$) the corresponding electron state at $\nu=n/(2mn+1)$
is incompressible, resulting in the FQHE.

In the QPH description, the quasielectron is assumed to have
intrinsic charge $-e^*$, local charge also $-e^*$, bosonic
exchange statistics, and fractional
anyon statistics $\theta$. Below we show, in several examples, that
this description also produces the same answer for the phases as
the mean-field approximation of Eq.~(\ref{phase}). We find
it convenient to use
a slightly different convention for the statistics of the
QPH-quasiparticles. We assume that they obey fermionic exchange
statistics and anyon statistics $\theta^*$ \cite {comm}, where
\begin{equation}
\theta^*=\theta-1=-\frac{2m}{2mn+1}\;\;.
\label{theta*}
\end{equation}
The total statistics, including
in a single phase factor both the exchange statistics and the phase
coming from the dynamical interaction, is $\theta$ in either
convention.

\underbar{One quasiparticle:} Consider a system with only one
quasielectron, i.e., only one composite fermion in an otherwise empty
quasi-LL, with the lower $n$ quasi-LL's fully occupied.
Let the quasielectron go around a closed loop in the counterclockwise
direction. The {\em average} phase associated with this path is
$2\pi B^*A/\phi_{0}$. In the QPH picture, the phase associated with
this path is $2\pi BA/\phi_{0}^*$ where
$$ \phi_{0}^*=\frac{hc}{e^*}=(2mn+1)\phi_{0}\;.$$ The two are equal since
$B^*=B/(2mn+1)$ for incompressible states (or, $eB^*=e^*B$).
The CF result would be obtained if the
Berry phase calculation of Ref. \cite {ASW} is interpreted
in terms of an effective magnetic field rather than a fractional
intrinsic charge.

\underbar{Two quasiparticles:}
Now consider two quasielectrons,
and let one go around a closed loop
in the counterclockwise direction. The {\em average} phase is
$2\pi BA/\phi_{0}-4m \pi \rho A$ when this loop does not enclose the
other quasielectron,
and $2\pi BA/\phi_{0}-4m \pi (\rho A+ e^*/e)$ when it does.
The difference between the two is $2\pi\theta^*$ with
\begin{equation}
\theta^*=-2m\frac{e^*}{e}\;\;,
\end{equation}
in agreement with the anyon statistics, Eq.~(\ref{theta*}),
of the QPH theory.
This is a ``derivation" of anyon statistics starting from the
CF theory. Note that the mean-field approximation of
averaging over positions of other composite fermions is
necessary for obtaining fractional statistics (which entails
non-analyticity); for any given configuration of other composite
fermions, the phase due to vortices is always a multiple of $2\pi$.

\underbar{Several quasiparticles:}
Away from the special filling factors, when there is a finite density of
quasiparticles, assumed to be uniformly distributed,
the average phase associated with a closed loop is
still given by  Eq.~(\ref{mfphase}) according to the CF theory.
This result is obtained in the QPH theory as follows. Consider
a magnetic field $B<B_{0}$ where $$B_{0}=\frac{2mn+1}{n}\rho\phi_{0}$$
corresponds to the FQHE state at $\nu_{0}=n/(2mn+1)$. The
quasielectron density per unit area at $B$ is given by
$$\rho_{q}= n\frac{(B_{0}-B)}{\phi_{0}}\;\;.$$
(This follows since each flux quantum
away from $B_{0}$ produces $n$ quasiparticles.) The average phase
associated with a closed loop is then given by
\begin{equation}
2\pi A (\frac{B}{\phi_{0}^*}+ \rho_{q}\theta^*)\;\;,
\end{equation}
which is the same as Eq.~(\ref{mfphase}).

\underbar{Resonant tunneling:}
Consider next a resonant tunneling situation in which a composite fermion
tunnels from one edge of the sample to the other through a path that
goes around a potential hill containing no electrons.
According to Eq.~(\ref{phase}), with $K=0$, the
phase associated with this path is simply the usual AB phase
\begin{equation}
2\pi\frac{BA}{\phi_{0}}\;\;,
\end{equation}
which implies that successive resonant tunneling
peaks are expected when the flux through
the electron-free region changes by $\phi_{0}$. This period was anticipated
theoretically \cite {Kivelson} and has also been observed
experimentally \cite {Goldman,Ford}. Note that the
charge deficiency of the electron-free region
is an integer multiple of the {\em local} charge $e^*$, as
observed experimentally \cite {Goldman}. In the
quasiparticle picture, if we assume that the quasiparticles, their
fractional charge and their fractional statistics are well defined at
the boundary of the FQHE state (which is
a non-trivial assumption due to the absence of a gap there),
and also recognize that the electron-free region is
made of $N_{q}=nAB/\phi_{0}$ quasiholes, then the phase is given by
\begin{equation}
2\pi\frac{ BA}{\phi_{0}^*}+2\pi N_{q}(-\theta^*)
\end{equation}
where $-\theta^*$ is the relative statistics of quasielectron
and quasihole. This also reduces to the previous equation.

\underbar{Spherical geometry:}
Interpretation by fractionally-charged fractional-statistics
quasiparticles becomes more complicated in spherical
geometry \cite {Haldane83}, where the electrons move on the surface of
a sphere under the influence of  a radial magnetic field.  In this
geometry, the electron system at flux $N_{\phi}$ is
mapped on to the CF system at flux
\begin{equation}
N_{\phi}^*=N_{\phi}-2m(N-1)\;\;,
\end{equation}
which reduces to Eq.~(\ref{phase}) in the limit of large $N$, with
$\nu=N/N_{\phi}$ and $\nu^*=N/N_{\phi}^*$.
This equation has been tested in numerical work on small systems,
which have shown a striking similarity between the
low-energy spectrum of interacting electrons at $N_{\phi}$ and that of
noninteracting fermions at $N_{\phi}^*$ in a broad range of filling
factors \cite {Dev}. For a single composite fermion in the $n$th
quasi-LL (with $n=1$ being the lowest quasi-LL),
the  phase associated with a closed path is
\begin{equation}
L^*\Omega\;,
\end{equation}
where $L^*=\frac{N_{\phi}^*}{2}+n-1$,
and $\Omega$ is the solid angle of the path.
In the quasiparticle picture, it would be natural to write
the analogous phase as \cite {Einar}
\begin{equation}
2\pi \frac{e^*}{e} \frac{\Omega}{4\pi} N_{\phi}\;\;,
\end{equation}
which differs from the previous equation by a term of order unity.
This term has been  interpreted as a new quantum number of the
quasiparticle, the ``spin" \cite {Einar}.
(It has nothing to do with the actual spin of the electron.
The electrons have been assumed to be spinless in this work, as
appropriate for a fully polarized system of electrons.)
The expression for the spin is, however, rather complex, and does not
satisfy any simple relation with the anyon statistics of the quasiparticles.
In particular, the quasiparticle and the quasihole spins are not equal
in magnitude. In agreement with an earlier suggestion \cite {Goldhaber},
Einarsson {\em et al.} \cite {Einar} have shown that
the part of the spin which is even under charge conjugation (i.e., the
average of the quasielectron and quasihole spins) is well defined and
satisfies the usual connection with the fractional statistics
$\theta^*$.

{\bf EFFECTIVE ELECTRIC FIELD}

An effective `renormalization' of the magnetic field implies an
effective `renormalization' of the electric field due to
the following argument. Assume a uniform electric field $E$ in the
plane of the electron system, so electrons move with velocity $v=cE/B$ in
a direction transverse to $E$. This result is independent of the
nature of the interaction-induced correlations, most easily
seen by going to a new frame of reference moving with velocity $v$,
where there is no electric field. Therefore, composite
fermions, which  experience an effective
magnetic field $B^*$ rather than $B$, must also experience an
effective electric field $E^*$, satisfying
\begin{equation}
\frac{E^*}{B^*}=\frac{E}{B}\;\;
\end{equation}
i.e.,
\begin{equation}
E^*=E-2m\rho\phi_{0}\frac{E}{B}\;\;.
\end{equation}
As in Eq.~(\ref{mfphase}), the second term on the right hand side
is induced by the vortices
carried by electrons,  as explained in Ref. \cite {ZHK}
in the context of the boson Chern-Simons theory. We repeat the
argument briefly. Consider the current  $e\rho v W$ through a cross section
of width $W$. Since each electron carries with
it $2m$ vortices, there is also a vortex current given by
$2m\phi_{0} \rho v W$. This, by Faraday's
law, leads  to a potential $2m\phi_{0} \rho v W/c$ across this
region of width $W$, thereby providing the second term in the above
equation. For the incompressible state at $\nu=n/(2mn+1)$, the
effective electric field is given by $E^*=(e^*/e)E$, since
$E^*/E=B^*/B=\nu/\nu^*= 1/(2mn+1)=e^*/e$.

The concept of effective electric field helps resolve an apparent
problem with the interaction energy of two quasiparticles within the
CF scheme. The CF picture for the quasielectron is somewhat reminiscent
of an electron in a dielectric, which is screened by electric dipoles
to produce a local charge $-e_{L}$.
The Coulomb energy of two electrons embedded in the
dielectric, at a distance $r$, is $e_{L}e/r$, which is equal to
the work done in bringing an electron from infinity to $r$ under the
influence of the electric field of the charge $-e_{L}$ object at
the origin. The coupling of the incoming electron with the electric field is
with the full charge $-e$, rather than the local charge $-e_{L}$.
The same is true of coupling with the magnetic field, as required
by the gauge invariance of electrodynamics. Thus, the screening which
reduces the local charge has
no effect on the coupling with the electromagnetic field, a
circumstance which is true much
more generally for electrons in complex media \cite {GK}.
Now let us consider the $\nu=n/(2mn+1)$ FQHE
state. The quasielectrons of this state experience  the magnetic
field $B$ with the full charge $-e$, as implicit in  Eq.~(\ref{phase}).
By gauge invariance, this can only be true if the coupling to an electric
field also entails the same charge $-e$. Naively, i.e, without taking
account of the electric field renormalization, this would suggest that
the interaction energy of two FQHE quasielectrons is
$e^*e/r$. (For the discussion here, we assume that
the FQHE system is in vacuum, rather than being at the interface of
two dielectrics.) This is in contradiction with the correct result,
known to be $(e^*)^2/r$ \cite {Laughlin83}. However, the
interaction energy with the effective electric field is
given by $(e^*/e)e^*e/r$, reproducing the correct result.

{\bf DISCUSSION}

It might seem surprising that the two approaches obtain the same
answers for the long range interactions, both electric and magnetic,
of quasiparticles, starting from two rather
different viewpoints. However, this happens because
the very assumption of incompressibility at
a fractional filling factor puts strong constraints on these
quantities. In fact, Su \cite {WPSu} has shown that the long-range
electromagnetic interactions of
the quasiparticles are determined {\em uniquely} for any given
incompressible FQHE state, provided one assumes that there is only
one type of quasiparticle, which carries the largest allowed local
charge.

The QPH theory is formulated in terms of the quasielectrons,
i.e., the composite fermions of the topmost quasi-LL. The CF theory
on the other hand treats all composite fermions in an equivalent
fashion. The full treatment in terms of composite
fermions of various quasi-LL's becomes essential in the limit of
large $n$. Here, the gap between neighboring quasi-LL's disappears
and  it is not meaningful to describe the dynamics in terms of
fractionally-charged fractional-statistics quasiparticles, but the CF
description may, in principle, continue to be valid. In an insightful
work, Halperin, Lee, and Read \cite {HLR} interpreted certain
experimental anomalies near $\nu=1/2$, where no FQHE is observed,
as a signature of the existence of a Fermi sea of composite
fermions. Several subsequent experiments have produced evidence for
composite fermions and their Fermi sea in the vicinity of
1/2 \cite {CFexpt}. Note that, unlike for the incompressible FQHE states,
the local charge of the CF-quasiparticle, and consequently also its anyon
statistics are not well defined for the  the compressible $\nu=1/2$
liquid, due to the absence of a gap \cite {GK}, but its
intrinsic charge $-e$ and fermion exchange statistics are
still sharp observables.  The observed cyclotron radius at $\nu=1/2$
is consistent with the existence of a Fermi surface of charge $-e$
fermions, and provides direct evidence
for the intrinsic quantum numbers of composite fermions.
The Shubnikov-de Haas analysis of Ref. \cite {Leadley}, and the
explanation of the thermopower measurements of Ref. \cite {Ying} also
assume charge $-e$ fermions.

In conclusion, we have analyzed two descriptions for the quasiparticles
of the FQHE, one in which they are viewed as  novel, topological
objects with intrinsic fractional electric charge
and intrinsic fractional statistics, and the other in which they are
pictured as composite fermions, i.e., as dressed electrons.
The CF theory has the pleasing aspect that the quasielectrons are the
same composite fermions as those in the lower quasi-LL's,
in other FQHE states, or even in the compressible Fermi liquid state.
Furthermore, it reveals that a quasielectron is simply an
electron dressed by the composite fermion medium,
and hence has the same intrinsic charge, spin and exchange statistics
as an electron, independent of the background state.
Its local charge, on the other hand, depends on the screening
properties of the background CF state. This seems
to be very much in the spirit of quasiparticle descriptions used for
many other interacting electron systems in condensed matter physics.
The CF theory thus unifies the view of quasiparticle
dynamics in and even beyond the FQHE regime.  The remarkable new
physics of the FQHE arises from the special nature of
correlations which provide additional phases, leading to
what may be a unique phenomenon, the dynamical
renormalization of the electromagnetic field.

We have benefited greatly from extensive discussions with Steven Kivelson.
Torbjorn Einarsson, Hans Hansson and Jon-Magne Leinaas also made
instructive comments.  This
work was supported in part by the National Science Foundation under
grant numbers PHY93-09888 (ASG) and  DMR93-18739 (JKJ).


\begin{thebibliography}{99}
\bibitem{comm} This would be the result of the Berry phase calculation
of Ref. \cite {ASW} if the wave function for two
quasiholes at $\eta$ and $\eta'$  is taken to be
$(\eta-\eta')\prod_{j}(z_{j}-\eta)(z_{j}-\eta')\chi_{_{L}}$, as opposed
to the one without the factor $(\eta-\eta')$.
\end{thebibliography}
\end{document}